\documentclass[%
 aip,
cp,  
 amsmath,amssymb,
 reprint,%
]{revtex4-2}

\usepackage{graphicx}
\usepackage{dcolumn}
\usepackage{bm}

\usepackage[utf8]{inputenc}
\usepackage[T1]{fontenc}
\usepackage{mathptmx}

\begin{document}

\title{Construction and Calibration of a Streaked Optical Spectrometer for Shock Temperature}

\author{Erik J. Davies}
\email[Corresponding author: ]{ejdavies@ucdavis.edu}
\author{Dylan K. Spaulding}
\author{Sarah T. Stewart}

\affiliation{Department of Earth and Planetary Science, University of California, Davis, 1 Shields Ave, Davis, CA, 95616, USA }

\begin{abstract}
Here we describe the implementation and calibration of a streaked visible spectrometer (SVS) for optical pyrometry and emission/absorption spectroscopy on light gas gun platforms in the UC Davis Shock Compression Laboratory. The diagnostic consists of an optical streak camera coupled to a spectrometer to provide temporally and spectrally-resolved records of visible emission from dynamically-compressed materials. Fiber optic coupling to the sample enables a small diagnostic footprint on the target face and flexibility of operation on multiple launch systems without the need for open optics. We present the details of calibration (time, wavelength and spectral radiance) for absolute temperature determination and present benchmark measurements of system performance.    
\end{abstract}

\maketitle

\section{INTRODUCTION}
Thermodynamic properties of dynamically-compressed materials are important to understand across a wide range of pressure-temperature phase space. The UC Davis Shock Compression Laboratory studies fundamental material properties governing planetary structure and formation; areas in which accurate thermodynamic knowledge is crucial and yet largely undocumented for the major constituents of rocky planets. High-quality temperature measurements are notoriously difficult in dynamic high-pressure experiments due to the inherently short timescales, uncertainty in optical properties such as emissivity, and other complications introduced by the experimental environment that can bias or pollute observations of thermal emission from the sample. Carefully executed experiments with reliable calibrations are therefore critical for a number of disciplines. 

Pyrometry diagnostics and calibration techniques have varied widely in the shock-compression community \cite{fat2015contributed, miller2007streaked}. While different approaches (e.g. discrete, multi-color vs spectrally integrated measurements) have unique strengths, a broadband, time-resolved diagnostic returns the most complete thermodynamic information about the sample and the evolution of its optical properties under compression. Diverse calibration sources and techniques are reported in the literature, from traditional, absolute methods which reference a known spectral radiance to the increasing popularity of quartz as an in-situ calibrant for relative temperature determination [e.g. \citep{root2018forsterite, millot2015shock}]. Regardless of the technique, it is imperative that the calibration be robust, insensitive to minor perturbations in the experiment, and ideally independent of secondary corrections to relate the experimental data to the calibrant.   

Here, we describe the construction and calibration of a fiber-coupled Streaked Visible Spectrometer (SVS) designed for optical pyrometry and emission/absorption spectroscopy with high sensitivity and wide spectral response. The system is versatile for a number of different types of measurements, has a small physical footprint on the sample, and is easily and robustly calibrated against a NIST-traceable standard. The system is sensitive to relatively low-intensity calibration sources, permitting direct comparison to experimental data without additional scaling, correction factors or other analytical transformations to relate observed thermal emission to the calibration source. We describe how the system response is determined and show examples of benchmark experiments to test the calibration and sensitivity over a range of timescales. While the system description is specific to the UC Davis SVS system, the calibration methods may apply to any similarly configured diagnostic. 

\section{SYSTEM DESCRIPTION}

\subsection{Streaked Visible Spectrometer}

\begin{figure}[h!]
\centering
\includegraphics[width=4in]{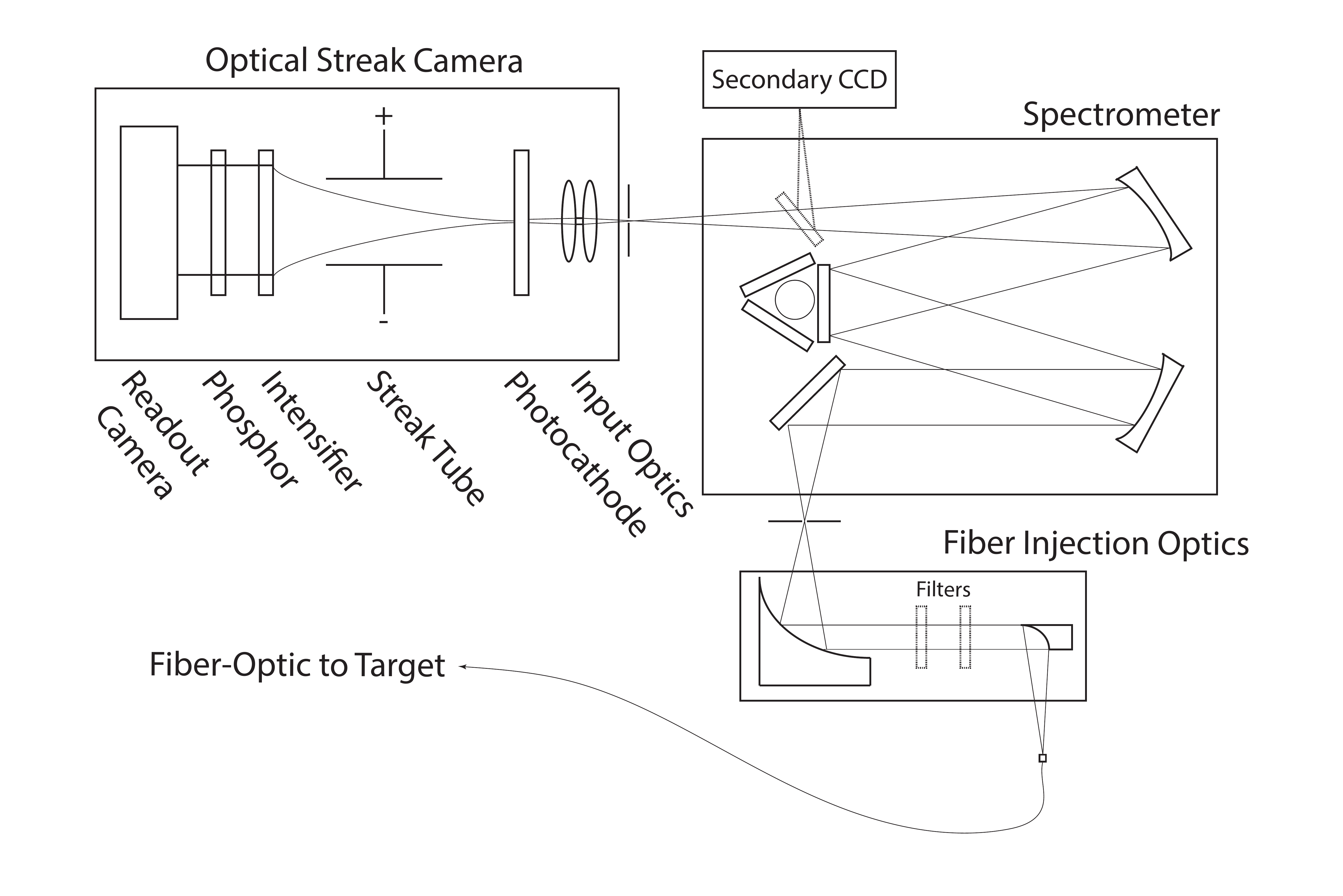}
\caption{Schematic of the streaked visible spectrometer including fiber probe from the experiment, injection optics, spectrometer, streak camera and readout. \label{fig:ShotSetup}}
\end{figure}

The streaked spectrometer consists of an Optronis SC20 streak camera \cite{OptronisProductLiterature}, coupled to a Princeton Instruments HRS300 spectrometer \cite{PrincetonInstruments} with protected Ag parabolic injection optics. A generalized schematic is given in Fig. \ref{fig:ShotSetup}. The system was conceived for maximum sensitivity to relatively low temperatures and components were therefore chosen for optimal sensitivity at the longest wavelengths allowed by the streak camera. 

Light is conveyed from the sample to the diagnostic via a single 300-$\mu$m, low-OH silica fiber. The target-facing end is a bare fiber so that the collection area on the sample is defined by the numerical aperture (0.39) and the standoff distance from the sample. Custom fiber rosettes are typically fabricated for simultaneous use of complimentary multi-color VIS/NIR pyrometry. A single, continuous fiber is used for the SVS, eliminating connection loss at the vacuum/air interface. The fiber remains connected to the spectrometer following calibration thereby ensuring that there is no alteration of the optical path between calibration and the science shot. The fiber output is imaged on the spectrometer entrance slit using two protected Ag parabolic reflectors which generate a short section of collimated beam for installation of neutral density or bandpass filters, as required.

The Princeton Instruments HRS300 spectrometer includes a variable entrance slit and a rotating turret with 3 grating selections of 150, 300 or 600 grooves/mm, allowing spectral resolution of $\sim$0.5~nm and variable central wavelength for detailed spectroscopy. A motorized mirror diverts the spectrometer output to either of two ports - one upon which a Thorlabs LC100 line CCD camera is mounted to check static throughput at the exit of the spectrometer without exposing the streak camera, and the second which diverts light directly to the slit of the streak camera.

\begin{figure}[h!]
\centering
\includegraphics[width=4in]{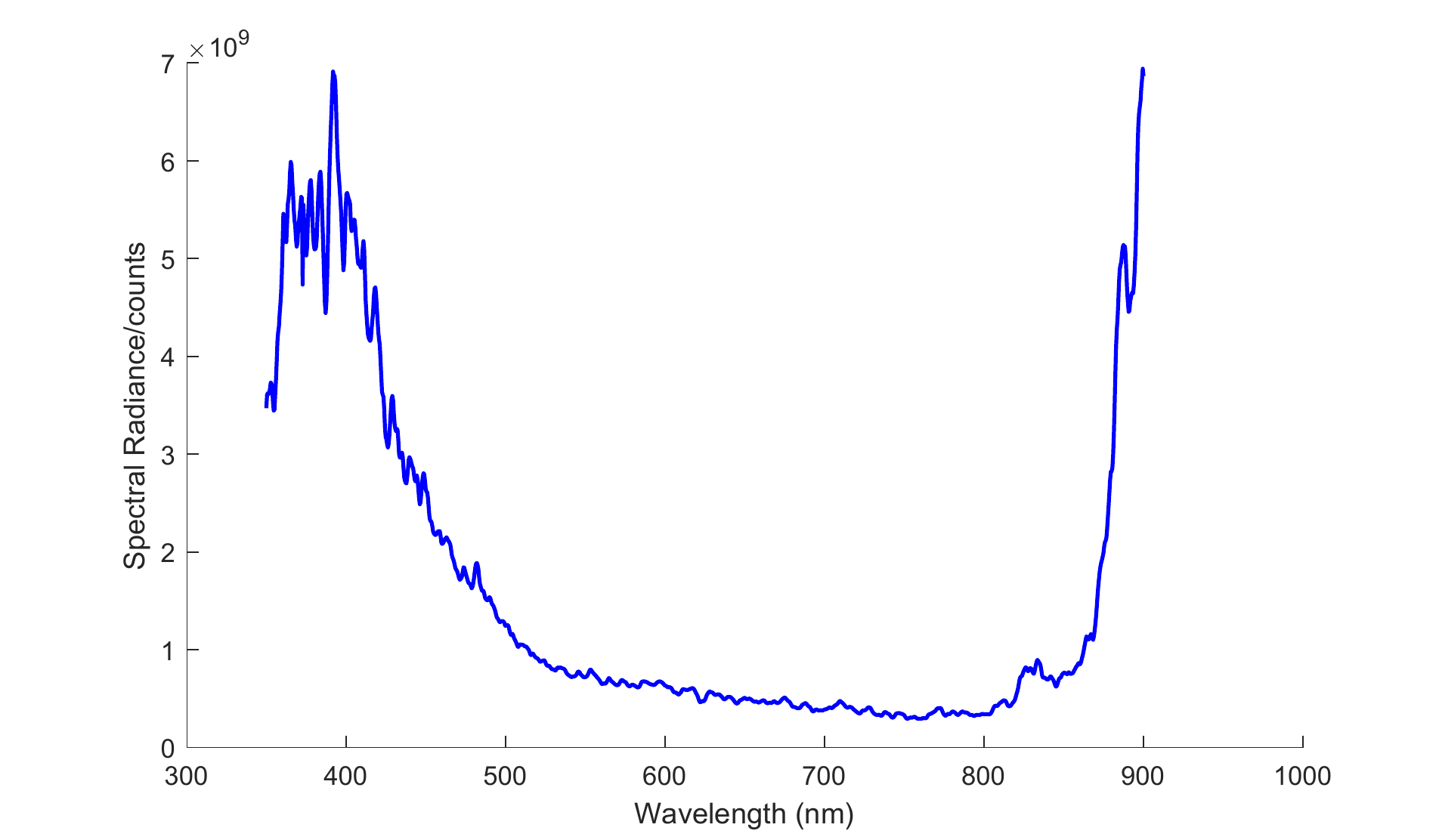}
\caption{Optical Response of the SVS ($C$, in Eq. \ref{eq:corr}), as determined with a tungsten-halogen calibration source via integrating sphere. The sharp rise above 875nm is due to the efficiency of the S25 photocathode in the streak camera and limits the lowest temperatures observable with the system. \label{fig:optic}}
\end{figure}

The system utilizes an Optronis SC20 optical streak camera equipped with a 35~mm $\times$ 4~mm S25 photocathode which provides the best long-wavelength sensitivity of the available options \cite{OptronisProductLiterature}. Two modular timing units allow sweep speeds from 112~ns to 700~$\mu$s total duration and a single-stage micro-channel plate permits variable gain. Readout is performed with a 2k $\times$ 2k, 16-bit, chilled CCD from Spectral Instruments Inc. \cite{SpectralInstruments}. The system response, shown in Fig. \ref{fig:optic}, spans $\sim$375 to 875~nm with optimum uniformity between $\sim$500~nm and 750~nm. 

\subsection{Calibration Hardware}

A variety of spectral radiance, blackbody, and illumination standards are used for absolute temperature calibration of science frames, characterization of system response, and measurement of transmission spectra for neutral density and bandpass filters. Calibration equipment is permanently located adjacent to the spectrometer to permit easy data acquisition from any source by moving a single fiber connection. Two additional USB spectrometers (StellarNet BLK-CXR) allow real-time signal comparison between detectors or quick diagnostic spectra during calibrations. 

The SVS system response is absolutely calibrated for dynamic pyrometry measurements using a Gooch and Housego OL455 tungsten-halogen lamp with a NIST-traceable spectral radiance curve. An additional Mikron M360 tuneable blackbody source ($\sim$873~K to 1373~K) or Optronics Labs OL550 tungsten ribbon lamp (also NIST traceable) are used for cross-comparison. Temporal calibration of streak images is performed using a Thorlabs NPL52B pulsed laser (450~nm, 1-10 MHz) which functions as an optical timing comb for in-situ characterization of streak linearity. Spectral calibration is carried out using an HgAr pencil lamp, which has emission lines spanning the spectral response of the overall system ($\sim$375-875~nm).

Finally, the SVS setup includes a laser-driven Xe plasma white light source (LDLS, Energetiq EQ99X). This high brightness source is used to characterize filters and may be used to correlate measured intensity at fast sweep speeds with absolute calibrations performed over longer durations (though this is not typically necessary for observations on $\mu$s timescales typical of light gas gun experiments). Tests are currently underway to apply it for sample illumination for dynamic broadband emissivity measurements.  

\section{CALIBRATIONS}

\subsection{Direct Calibration}

To determine absolute temperature in any pyrometry system, the measured signal must be related to a source for which the spectral radiance (emission, measured in W~sr$^{-1}$~m$^{-2}$~nm$^{-1}$) is known as a function of wavelength. The relationship between the known spectral radiance and the measured signal defines the system response of the diagnostic. The system response is measured prior to every experiment to include variations in fiber-optic probe efficiency and user-selected choices such as camera settings and attenuation. The system response can be determined in a direct manner for the setup described above because it is sensitive enough to detect the calibration sources on experimental timescales, eliminating the need for additional corrections due to integration time. Furthermore, distant extrapolation between calibrated sources and experimental conditions are not required for the temperatures typically generated in light gas gun plate impact experiments. Calibration without signal level extrapolation is often not possible for other HED experiments at more extreme conditions. 

Consider an ideal greybody, defined by Planck's law, at some known temperature. The spectral radiance is defined as
\begin{equation}
    S_\lambda (\lambda,T) = \epsilon \frac{2h c^2}{\lambda^5} \frac{1}{e^{\frac{h c}{\lambda k_B T}}-1},
    \label{eq:planck}
\end{equation}
where $\epsilon$ is the emissivity, $\lambda$ is the wavelength, $T$ is temperature, $h$ is the Planck constant, $c$ is the speed of light, and $k_B$ is the Boltzmann constant. The number of counts recorded by the detector observing the greybody at temperature $T$ is simply given by 
\begin{equation}
    S_{Meas} = \frac{S_{real}}{C},
    \label{eq:corr}
\end{equation}
where $S_{Meas}$ is the measured number of counts, $S_{real}$ is the known emission spectrum, and $C$ is the optical response of the system. The optical response is then applied to the measured experiment image to generate a calibrated image. This simple correction is based on three assumptions: first, that the emitting surface is Lambertian (i.e., that the apparent radiance is independent of viewing angle); second, that the acceptance angle of the fiber does not include light from edge effects or other heterogeneities; and third, that emissivity, $\epsilon$, is wavelength independent. This latter assumption, though commonly applied, is rarely if ever true; however, the advantage of a spectrally resolved measurement is that the deviation of the raw data from an ideal greybody can be observed and additional forms of $\epsilon$ that are not constant and are dependent on wavelength can be fit. Furthermore, by illuminating the sample and measuring dynamic reflectivity, the same diagnostic can be used to measure the wavelength dependence of $\epsilon$ under experimental conditions.

Corrections to the acquired image are minimal and include subtraction of dark current (determined by taking an acquisition with no incident light on the streak camera), and temporal and spectral image distortion corrections to correct for slight non-linearities induced by the streak tube. These are done using the in-situ timing comb and a separate pre-shot acquisition of the emission lines from the HgAr pencil lamp. For sources beyond the dynamic range of the diagnostic, neutral density (ND) filters may be used and must be accounted for multiplicatively. Each ND filter is calibrated individually to generate a wavelength-dependent correction.

While not applied here, an `indirect' calibration technique can be used in cases where the integration time of a particular experiment is too short for the diagnostic to detect calibration sources on the same timescales. In this case, the LDLS white light source (described above) may be used to correlate measured intensities for different camera sweep rates. For the system described above, we have empirically determined that the relationship between measured counts and sweep rate is linear.

\subsection{Sources of Uncertainty}

Several sources of uncertainty affect the accuracy of pyrometry experiments. Uncertainty in the spectral radiance of the Gooch and Housego OL455 tungsten-halogen source and the Mikron blackbody source is about 1\%. The blackbody oven is assumed to have an emissivity of 1. ND filter corrections are also assumed to have an uncertainty of $\sim$1\% due to variations in the stability of the light source used to characterize them and the potential for degradation over time.

For every acquired image, the uncertainty is assumed to scale as the $\sqrt{S_{\rm meas}}$ for each pixel. The random uncertainty in counts is propagated through every image operation (e.g., dark subtraction) and is on the order of 1\% for most calibrated images. Variation in counts during the duration of interest in the experiment also contributes to uncertainty in the final temperature determination. Averages are taken over line-outs with a standard deviation that depends on the temporal width of the region of interest. Longer averages will therefore generate more precise measurements barring rapid changes in temperature over that interval. As described above, the system has been designed to eliminate the potential for changes in the optical path (hence system response) since these cannot be quantified if they occur after the calibration. All reported uncertainties are 1-sigma in this work.

\section{PROOF OF CONCEPT: QUARTZ STANDARD AND BLACKBODY CROSS REFERENCE}

Two benchmark examples are shown here to demonstrate the performance of the SVS diagnostic. Both examples are fit to equation \ref{eq:planck} using the method of least-squares, with either $T$ or $T$ and $\epsilon$ as fitting parameters, as discussed below. 

The first example was conducted statically as a benchtop experiment using a LAND blackbody oven (on loan from Lawrence Livermore National Laboratory) at 1453 K as a calibration source. The UC Davis Mikron blackbody oven, with an internal temperature of 1349 K was treated as an `unknown' sample. Both oven set points were verified separately using a Minolta Cyclops 52 handheld pyrometer. A sweep rate of 25 $\mu$s/mm was used on the camera to enable sensitivity below 1500 K. A fit averaged over 350 $\mu$s yields a calculated temperature of T=1348$\pm$37 K. In this case, the calibration ovens are known to have a constant emissivity very near unity, so $T$ is the only fitted parameter and $\epsilon$ is assumed constant. At timescales representative of most light gas gun experiments (several $\mu$s), the lowest observable temperatures are $\sim$2000~K due to the shorter acquisition time. 
\begin{figure}[ht]
\centering
\includegraphics[width=6.5in]{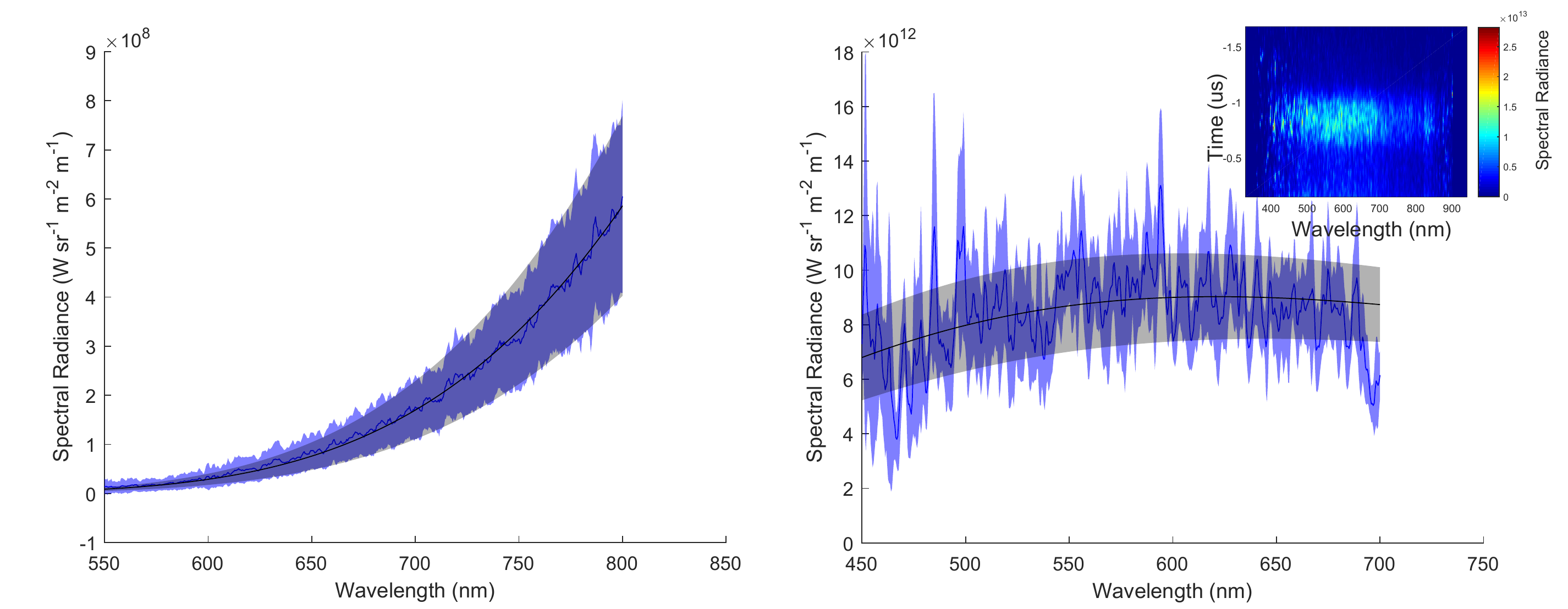}
\caption{ (Left) Observed spectral radiance for the MIKRON blackbody source (blue with magenta uncertainty bands) and the Plank fit (black with grey uncertainty bands) corresponding to a calculated temperature of 1348$\pm$37 K. 
(Right) Spectral radiance spectrum of the quartz shock front. The inset shows the experiment region of interest with spectral radiance given by the hot map. The spectrum from which temperature is calculated is a line out of the bright region of the map.
\label{fig:Oven}}
\end{figure}

For the second example, a plate impact experiment was conducted on the UC Davis two-stage light gas gun. The flyer and driver were 304 Stainless steel with an impact velocity of 5400$\pm\ 65$ m/s. The sample was z-cut $\alpha$-quartz, with a measured shock velocity from sample transit time measurements of 8.46$\pm.015$ km/s, corresponding to a pressure of 86.68$\pm$.68~GPa \cite{knudson2013adiabatic}. A Gooch and Housego OL455 tungsten-halogen lamp was used as the calibration source. The sweep rate was 1 $\mu$s/mm and an ND3 filter was used to attenuate the signal. The spectrum from the shocked quartz was averaged over $\sim$1~$\mu$s and is shown in Fig.~\ref{fig:Oven} (right). The shorter observation duration ($\sim$1/28th of the entire streak length) results in lower signal-to-noise in the data. Faster sweep rates at higher time resolution are needed to average random uncertainty of the experiment.

We present two fitting cases to illustrate the capabilities and representative uncertainties of this diagnostic. The first case lets $T$ and $\epsilon$ be free parameters in fitting equation \ref{eq:planck} and the second only fits $T$. In both cases, $\epsilon$ is assumed to be independent of wavelength. For quartz shocked to 106.5 GPa, the emission spectrum appears as a graybody and the unshocked quartz is totally transparent \cite{luo2004shock}. The experiment here is also of sufficiently low pressure that the emissivity of silica during the shock approaches 1 \cite{hicks2006dissociation, millot2015shock}. Over the wavelengths measured in this work, emissivity dependence on wavelength is not expected to be appreciable.

In the first case, the measured temperature is 4716$\pm$210~K, with $\epsilon=0.864\pm0.091$. The emissivity in this case is smaller than the measured $\epsilon$ in \cite{hicks2006dissociation}; however,uncertainty improves with faster sweep speeds, which improves signal-to-noise ratios (and temporal resolution) by spreading the data over a larger fraction of the CCD. For the second case, at the shock pressure of this experiment, the reflectivity is negligible \cite{hicks2006dissociation, millot2015shock} and we assume reflecticivity $R=0$ and $\epsilon=1$. In this case, the measured temperature is 4661$\pm$172~K. Both cases give temperatures that are consistent with previously measured quartz temperatures at 86$\pm$1~GPa and 4860$\pm$150~K \cite{lyzenga1980shock}. 

\section{CONCLUSION}
We have built and commissioned a streaked visible spectrometer for optical pyrometry and emission/absorption spectroscopy in the UC Davis Shock Compression Laboratory. Calibration schemes have been cross-referenced using multiple sources, and we have shown that the system enables observation of relatively low dynamic temperatures with high fidelity. With a small fiber-optic footprint on the sample, it may be fielded in parallel with a number of other optical diagnostics, including complimentary, multi-channel visible and NIR pyrometry and velocimetry. Additional applications could extend to Raman spectroscopy, dynamic reflectivity and other studies of high-pressure optical and electronic properties.     


\section{ACKNOWLEDGMENTS}
This work was supported by NASA grant NNX15AH54G, NASA grant NNX16AP35H, LLNL contract No.\ B617085, and UC Office of the President grant LFR-17-449059.


\bibliography{bibtex}%

\end{document}